# Determination of characteristic relaxation times and their significance in glassy Anderson insulators

T. Grenet and J. Delahaye

*Institut Néel, CNRS & Université Joseph Fourier, BP 166, F-38042 Grenoble Cédex 9*

**Abstract** – We revisit the field effect procedure used to characterise the slow dynamics of glassy Anderson insulators. It is shown that in the slowest systems the procedure fails and the "characteristic" time values extracted are not intrinsic but determined by the experimental procedure itself. In other cases (like lightly doped indium oxide) qualitative indications about the dynamics might be obtained, however the times extracted cannot be seen as characteristic relaxation times of the system in any simple manner, and more complete experiments are necessary. Implications regarding the effect of carrier concentration on the emergence of glassiness are briefly outlined.

**Introduction** – Typical glassy phenomena such as long time relaxations, memory of history and physical ageing exist at low temperature in the conductance field effect anomaly observed in thin film disordered insulators such as micro-crystalline and amorphous indium oxide[1] and granular metals[2,3]. Recent experiments on discontinuous Ni films, consisting of ferromagnetic metallic islands, have recently shown that the slow dynamics can be significantly altered by the magnetic field[4].

These phenomena are believed to be the signatures of a Coulomb Glass state of the charge carriers, which was first suggested more than two decades ago[5] and may result from the strong coulomb interactions between the poorly screened localized carriers of these highly disordered insulators.

Although a large amount of theoretical work has been performed recently there is yet no established precise scenario for the appearance of the glassy behaviour. One of the most important open questions is why the typical experimental signature (slow logarithmic growth of a field effect anomaly after a quench) has not yet been observed in standard doped semiconductors, the archetype of Anderson insulators. One suggestion is that high enough carrier concentrations are necessary, which cannot be realized in standard doped semiconductors without getting into the metallic state. This suggestion was based on a series of experiments on indium oxide samples with various oxygen deficiencies and thus various electron concentrations. It was shown that a characteristic relaxation time $\tau$ defined in a way we shall describe and discuss below, increases drastically when the electron concentration exceeds a few $10^{19}$ cm$^{-3}$ (Ref. 6) and then saturates above a few $10^{20}$ cm$^{-3}$.

The experimental implementation of $\tau$ is now used to characterize and compare the dynamics of other glassy disordered insulators. However in the present paper we shall stress that the definition used for $\tau$ can be problematic. In particular we show with new measurements that in "highly doped" samples its value is determined by the experimental procedure so that the apparent saturation of the dynamics is an artefact. In "lightly doped" samples different values for $\tau$ are observed however their meaning is not straightforward and more complete experiments are necessary to conclude about the nature of the changes observed in the dynamics. Some implications of these findings will be briefly discussed.

**Experimental definition of $\tau$** – One general property of the systems under interest is that after a quench from high to liquid He temperature, their electrical conductance is not constant but decreases logarithmically with time, with no reported saturation up to several days or weeks of measurement. As is well known this corresponds to a $1/\tau_i$ effective time distribution, *which has no characteristic value*[7]. It is also very difficult to directly compare the dynamics of different samples, or of one given sample at different temperatures, as the physics of the pre-factor determining the amplitude of the relaxations (the slope in a logarithmic plot) is not yet established. In an attempt to circumvent these difficulties, a "two dip" protocol was used to define a "characteristic time" $\tau$ of indium oxide samples[6]. It is based on the slow dynamics of the field effect anomaly observed in all the glassy systems of interest here and is recalled in Fig. 1.

For such experiments, the samples consist of MOSFET-like devices with a thin film of the disordered insulator constituting the (poorly) conducting channel. After a sample has been quench-cooled at a given gate voltage $V_g = V_{g1}$ and has been allowed to relax for a given time $t_{w1}$, any sudden change of $V_g$ away from $V_{g1}$ re-excites the system which was slowly approaching equilibrium, and restores the higher off equilibrium conductivity it had before relaxing. However the system keeps a memory of its relaxation under $V_{g1}$. Thus a fast $V_g$ scan reveals a conductivity "dip" as seen in the upper curve of figure 1-b.

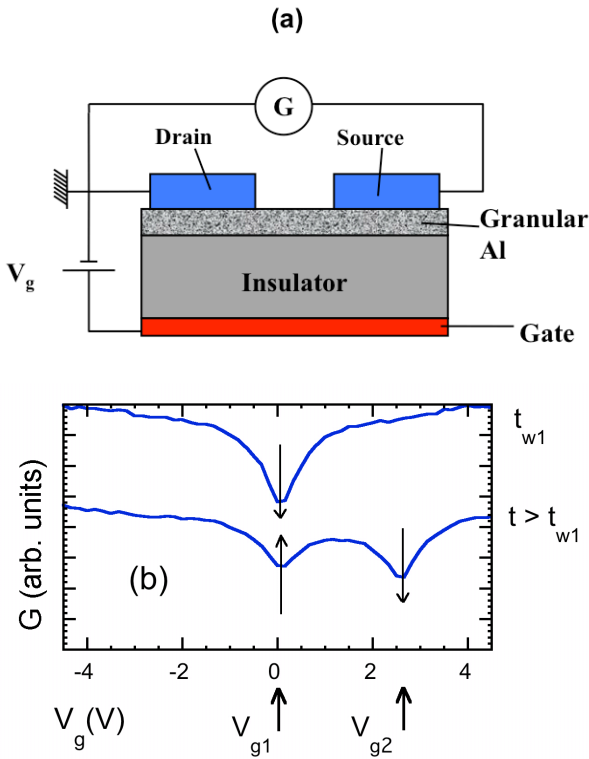

Figure 1: (a) Scheme of a MOSFET device used to measure the field effect dip in granular aluminium. (b) Illustration of the two dip protocol (see text). Upper curve: field effect dip formed at $V_{g1}$ during the first step of the protocol (sample cooled and maintained under $V_{g1}$ for $t_{w1}$). Lower curve: fading $V_{g1}$ dip and growing $V_{g2}$ dip during the second step of the protocol (sample maintained under $V_{g2}$).

If the gate voltage is set at another value $V_{g2}$, subsequent fast $V_g$ scans (separated by "long" periods of time with $V_g=V_{g2}$) show a fading dip at $V_{g1}$ and a slowly emerging one at $V_{g2}$, indicating the system's relaxation to a new equilibrium state under $V_g = V_{g2}$ (lower curve of figure 1-b). Using this "two-dip" protocol a characteristic time $\tau$ was defined as the time to be spent at $V_{g2}$ until the two dips are of equal amplitude. It was determined by regularly performing fast $V_g$ scans in order to detect when the two cusp amplitudes have crossed. The first step was supposedly performed during a long enough time $t_{w1}$ (12 hours) so that the system "has equilibrated" before $V_g$ is changed to $V_{g2}$.

A systematic study showed that monitoring the electron concentration by changing the oxygen deficiency of indium oxide films has a dramatic effect on the value of $\tau$: while it is of the order of $10^3$ secs for samples with $n_e > 10^{20}$ cm$^{-3}$ it seems to very abruptly decrease when $n_e < 10^{20}$ cm$^{-3}$ (see Fig. 3 of Ref. 8). This is believed to be an important clue of why the gate voltage dip and its glassy behaviour is not observed in standard doped semiconductors: these have too low carrier concentrations in their insulating state to have macroscopic relaxation times.

However we now wish to discuss difficulties associated with the definition and interpretation of $\tau$.

**"Highly" doped indium oxide and granular Al** – We need to distinguish the cases of "lightly" and "highly" doped indium oxide samples. We first discuss the case of the "highly" doped ones, as well as of granular aluminium. We associate these two materials in the discussion because they share the same glassy phenomenology. First we recall that the width of the field effect dip in indium oxide strongly increases with the electron concentration[6], and that the one observed in granular aluminium (when expressed in gate field) is the same as that of the most "highly" doped indium oxide samples[2]. Moreover when the protocol described above is applied to granular aluminium based MOSFETs a similar value for $\tau$ is obtained as with "highly" doped indium oxide.

If the systems under study were fully equilibrated during the first step of the protocol (i.e. if $t_{w1}$ was larger than the largest relaxation time of the systems) then the time $\tau$ obtained from the "two dip" protocol would indeed characterise the systems dynamics[9]. But such is not the case. Both highly doped indium oxide and granular aluminium, following a quench, display logarithmic relaxations for durations far exceeding the $t_{w1}$ used in the two-dip protocol. Thus at $t_{w1}$ the samples are generally *not equilibrated* and are *still logarithmically relaxing*. One may think this is unimportant for all practical purposes as long as the residual relaxation is very slow (logarithmic), but this turns out to be uncorrect[10]. Suppose that in the first step the first dip amplitude grows like $A\mathrm{Ln}(t/\tau_{micro})$ then its amplitude at time $t_{w1}$ is $A\ln(t_{w1}/\tau_{micro})$. Here $\tau_{micro}$ is the minimum relaxation time of the system. In the second step of the protocol, the second dip at $V_{g2}$ also grows like $A\ln(t'/\tau_{micro})$ while the first one at $V_{g1}$ decreases like $A(\mathrm{Ln}(t_{w1}/\tau_{micro})-\mathrm{Ln}(t'/\tau_{micro}))$ where $t'$ is the time spent in the second step. Equating the two amplitudes gives:

$$t' = \tau = \sqrt{t_{w1}\tau_{micro}}$$

This simple derivation assumes that the measurement itself ($V_g$ scans) is instantaneous and has no influence on the dips. This is of course not the case and since a typical measurement time is the duration $t_{scan}$ of a $V_g$ scan one may rather expect (see appendix for more justification):

$$t' = \tau \approx \sqrt{t_{w1} t_{scan}} \quad (1)$$

Thus the measured $\tau$ should be determined by the experimental protocol and not by the sole systems dynamics.

We performed new measurements on a granular aluminium film with $R(4K) = 20$ G$\Omega$ in order to check this. In Fig. 2 we show the $t_{w1}$ dependence of $\tau$. One clearly sees the square root dependence of $\tau$ with $t_{w1}$ up to $t_{w1}=2.59 \ 10^5$ s at least (three days).

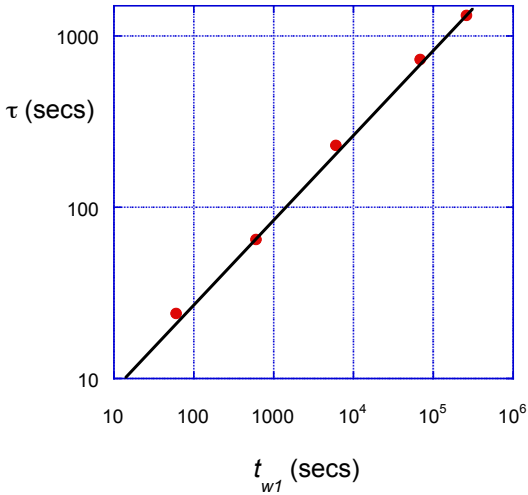

Figure 2: $t_{w1}$ dependence of $\tau$ in a granular aluminium thin film. $V_g$ scans are performed each minute. Here a $V_g$ scan consists of three conductance measurements performed at $V_g$=-5 V (baseline conductance), $V_{g1}$=0 V and $V_{g2}$=+5V. The straight line has slope ½ indicating a square-root dependence.

In Figure 3 we show the $t_{scan}$ dependence of $\tau$ for the same sample. One again sees a clear dependence, very close to a square-root one.

This establishes that in granular aluminium *the value of the two-dip "characteristic" time is determined by the experimental protocol and not by the systems dynamics.*

One wonders whether such an artefact is also present with "highly" doped indium oxide. This should be tested directly by future experiments but we tend to think the answer is yes. Indeed the value estimated using equation (1) (with $t_{scan}$=100-200 secs and $t_{w1}$=12 hours[6]) gives $\tau \approx$ 2-3 $10^3$ secs, very close to the experimental values (1-2 $10^3$ secs).

Thus it is highly probable that the constant value found for $\tau$ in the "highly doped" indium oxide samples (with $n_e$>$10^{20}$ cm$^{-3}$) is also solely determined by the experimental protocol. It does not indicate that the dynamics (which may be characterized by e.g. the largest relaxation times of the systems) is constant. The mere indication it gives is that the maximum relaxation times are larger than the experimental $t_{w1}$, and that the relaxations are all close to logarithmic with nearly the same prefactor A (as assumed in our derivation of equation (1) ).

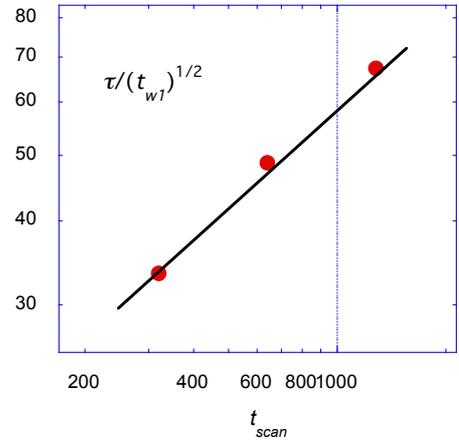

Figure 3: $t_{scan}$ dependence of $\tau$ in the same granular aluminium thin film as in Fig. 2. For these measurements the $V_g$ scans consisted of 31 measurements equally spaced between $V_g$=-5V and $V_g$=+5V. The different scan durations correspond to different single measurement durations (stays at the measured $V_g$s). Since for the three values of $t_{scan}$ the $t_{w1}$s were slightly different we plot $t/t_{w1}^{1/2}$ versus $t_{scan}$. The straight line has slope ½ indicating a square-root dependence.

**"Lightly" doped indium oxide -** We now wish to discuss the more interesting case of indium oxide with $n_e$<$10^{20}$ cm$^{-3}$ for which the "two dip" $\tau$ is significantly different (smaller) than equation (1). This may mean either that:
- i) the maximum relaxation times are smaller than the experimental times

or that
- ii) the dips relaxations are not symmetrical (unequal prefactors) or are not simply logarithmic

The published data give clues about what is happening. In Fig. 2 of Ref.11, reproduced here in Fig. 4, the two dips evolutions are shown for two indium oxide samples, a "highly doped" one (with $n_e$ = 5 $10^{20}$ cm$^{-3}$ and $\tau$ = 2-3 $10^3$ secs) and a "lightly doped" one (with $n_e$ = $10^{20}$ cm$^{-3}$ and $\tau$ = 50-60 secs). One immediately sees that hypothesis i) is not correct as the two dips amplitudes in the "lightly doped" sample, very much like in the "highly doped" one, are seen to relax with no clear sign of saturation up to times of $10^5$ secs, which is larger than the $t_{w1}$ used ! Surprisingly hypothesis ii) also seems to be contradicted by the data. As emphasized in the text, within the experimental accuracy the relaxations are logarithmic in time for both samples, with nearly symmetrical evolutions for each sample.

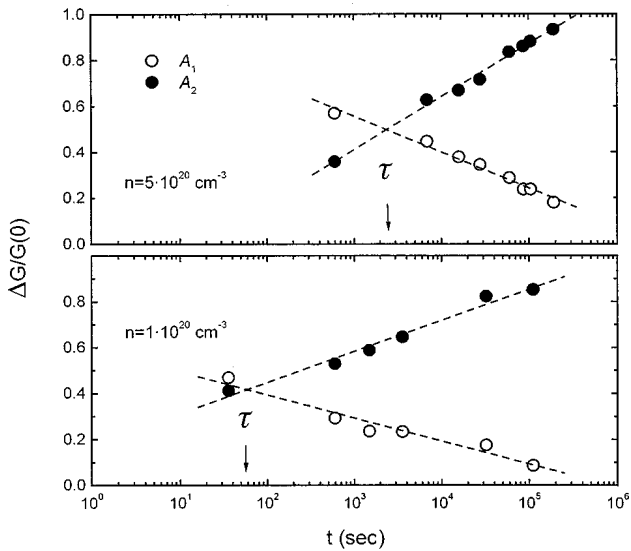

Fig. 4: time evolutions of the $V_{g1}$ and $V_{g2}$ dips during the second step of the "two dip protocol" for a "highly" (top) and a "lightly" (bottom) indium oxide sample (figure reprinted from Ref. 11, A. Vaknin, Z. Ovadyahu and M. Pollak, Phys. Rev. B **61**, 6692 (2000) ).

Moreover the slopes are smaller for the "lightly doped" sample than for the "highly doped" one. Accepting that one can compare the slow dynamics of samples with similar resistances at the same temperature, by comparing the slopes of the logarithmic relaxations of $\Delta G/G$, then one would conclude that the "lightly doped" sample (with the shortest $\tau$) is the slower one !

The resolution of this paradox may come from the *short time response* after the $V_g$ jump from $V_{g1}$ to $V_{g2}$, which seems to be different in the "lightly" and "highly" doped systems. Judging from the starting points of the curves, in the "lightly doped" sample the first dip seems to be quite rapidly erased after the gate voltage has been switched from $V_{g1}$ to $V_{g2}$ (it is reduced to a bit more than 40% of its initial value after 30 secs) while the effect is smaller or slower in the "highly doped" one (60% of its initial value after 600 secs). Similarly the growth curve of the second dip starts from a higher value in the lightly doped sample (40% of amplitude at 30-40 secs) than in the highly doped one (only 35% at 600 secs).

Thus it may be that the difference observed between highly and lightly doped samples comes from a "fast" response following $V_g$ jumps, at times too short to be investigated by the experimental procedure. "Lightly" doped systems would then have a relaxation time distribution with a larger weight at short times and a smaller one in the long time (logarithmic) region, as compared to the "highly" doped ones.

Unfortunately this simple conclusion cannot be taken for granted yet from the existing data, because what happens in the short time regime is not measured. It is desirable to find ways to investigate this regime.

Moreover one may wonder if the experimental procedure chosen influences the results for the lightly doped systems as well. We first note that the "two dip" protocol is clear if $t_{scan}$ is much smaller than the waiting time after the first scan and between successive scans. But this condition is not met with the "lightly" doped system so that in that case the effective waiting times for the two dips are not well defined. Second one may wonder if phenomena due to the $V_g$ jump itself (from $V_{g1}$ to $V_{g2}$) influence the results. Indeed it was already reported that $V_g$ jumps can rapidly destroy the first dip memory in indium oxide (see Fig. 13 and 14 of Ref.12), and there are indications that the effect is more prominent for "lightly doped" samples than for "highly doped" ones[13]. Although at that time no connection was made with the "two dip" characteristic time, this phenomenon may influence the $\tau$ values obtained for the lightly doped indium oxide samples.

**Implications** – The present analysis has several potentially important implications for the understanding of glassiness in Anderson insulators.

First the dependence of the dynamics on the carrier concentration is probably not the one suggested by the curve $\tau$ ($n_e$) accepted up to now, at least the saturation above a "critical" concentration around $10^{20}$ cm$^{-3}$ is not real. The sole $\tau$ values bring either misleading or far too incomplete information about the systems dynamics. In order to discuss two dip experiments, one has to compare the full relaxation curves, including the first step relaxation before $t_{w1}$.

Second the non-observation of any glassy field effect dip in "standard" semiconductors (like e.g. doped silicon) was up to now explained relying on the $\tau$ data suggesting a very steep decrease of $\tau$ when $n_e$ becomes smaller than $10^{20}$ cm$^{-3}$. For this reason doped semiconductors (of carrier concentration less than $10^{19}$ cm$^{-3}$ in their insulating regime) were believed to have un-measurably fast relaxation times. This explanation has to be reconsidered or refined in the light of our findings. If $V_g$ changes indeed have strong detrimental effects on the memory, then in principle no gate voltage dip can be measured in these systems using $V_g$ scans. It might then be more interesting to search for slow relaxations after the quench at constant gate voltage.

Finally experiments aimed at comparing the dynamics of different systems using the "two dip" protocol should be considered with much care. One must remember that by construction the "two dip" protocol *does not provide an absolute measurement of a given system's dynamics, but only a comparison of its dynamics in the first and second steps, i.e. before and after the gate voltage jump*[14]. Thus it may rather be used to study the effect of an external parameter like e.g. temperature or magnetic field on the dynamics of one given system, provided one changes the external parameter value at the same time as the $V_g$ jump (see e.g. the case of T changes in Ref.2), and provided possible effects of the $V_g$ jump itself on the dynamics can be dealt with. An even better way is probably to study whether the growth of a given dip, submitted to different values of the external parameter during succeeding time intervals, is cumulative or not (like in Fig. 10 of Ref. 15). This procedure seems to be the most reliable envisaged so far.

**Conclusion** – We have revisited the main experimental procedure used so far to characterise the dynamics of the glassy disordered insulators. We have shown that the "characteristic" times $\tau$ extracted from the "two dip" protocol can be misleading. For systems behaving like "highly doped" indium oxide (including granular aluminium), values obtained for $\tau$ are useless as they are determined by the experimental procedure (equation (1) ) and not by the systems properties. For the "lightly doped" indium oxide case, the small values of $\tau$ seem to be related to short time effects following the gate voltage jump in the "two dip" protocol. The "characteristic" time $\tau$ alone does not constitute a good indicator of the systems dynamics. Indeed it is observed that both large and small $\tau$ valued indium oxide samples have slow logarithmic relaxations extending to very large times (up to $10^5$ secs at least).

In general in order to understand departures of $\tau$ from the value given by equation (1) it is important to compare the full relaxation curves of all the steps of the experiments protocols.

In the case of "lightly doped" indium oxide samples, future experiments seem necessary to better understand the short time responses. Relaxation measurements after a $V_g$ jump without performing complete $V_g$ scans (which take time and prevent the short times to be explored), as well as relaxation measurements after a quench (short time relaxation without any $V_g$ jump) should bring many interesting results.

The common explanation for the non-observation of any gate voltage dip in standard doped semiconductors might also be revisited in the light of the present remarks.

**Acknowledgment** – We very acknowledge financial support from the Université Joseph Fourier (SMINGUE 2010 program) as well as from the Region Rhône-Alpes (CIBLE 2010 program). Discussions about this topic with Ariel Amir, Stefano Borini and Zvi Ovadyahu are kindly acknowledged.

**Appendix 1** – Here we justify formula (1). Suppose a sample is quench-cooled at $t$=0, relaxed under $V_{g1}$ for $t_{w1}$ and then relaxed for a time $t_{w2}$ under $V_{g2}$. Then a $V_g$ scan is performed and for simplicity we suppose that the value of $t_{w2}$ was wisely chosen so that the two dips are found of equal amplitude (i.e. $t_{w2}$=$\tau$). We choose (like in our experiments) that the $V_g$ scan is performed in the direction of increasing gate voltage. In general as long as $V_g$ is nearly equal to a given value $V_{gi}$ the $i$th dip amplitude (centered on $V_{gi}$) increases with time while the $j$th dip amplitude (if any) decreases. Using the "superposition principle" already discussed in Ref. 3 the time evolution of the dips amplitudes is a sum of positive and negative logarithms. The evolutions of the two dips amplitudes may be schematized as in Fig. 5.

Their expressions are:

$$\Delta G(V_{g1}) = A\{Ln(t) - Ln(t-t_{w1}) + Ln(t-t_{w1} - \tau - (t_{scan} - t_{scandip})/2)\}$$

$$\Delta G(V_{g2}) = A\begin{Bmatrix} Ln(t-t_{w1}) - Ln(t-t_{w1} - \tau) \\ + Ln(t-t_{w1} - \tau - t_{scan} + t_{scandip}/2) \end{Bmatrix}$$

where $t_{scandip}$ is the time it takes to scan a dip width. The $\tau_{micro}$ denominators have been omitted in the logarithms as they cancel in the calculation. Equating the two dip amplitudes *each at the time when they are measured*, gives a relation which solution is:

$$\tau = \tfrac{1}{2}\sqrt{(t_{scan}/2)^2 + 4t_{scan}t_{w1}} - t_{scan}/4$$

which is very close to equation (1) as long as $t_{scan} \ll t_{w1}$.

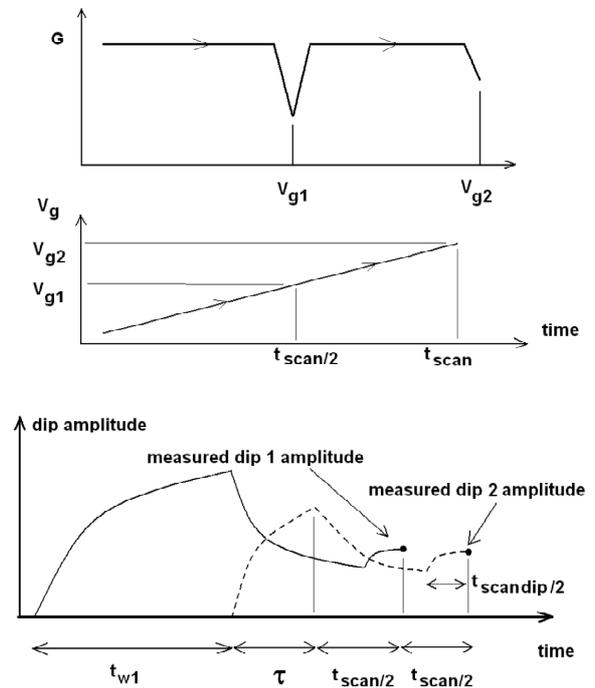

Figure 5: <u>Top and middle</u>: scheme of a $V_g$ scan during the second step of the two-dip protocol. <u>Bottom</u>: time evolution of the dips amplitudes in the two-dip protocol with only one $V_g$ scan started at time $\tau$ after $t_{w1}$. Time intervals are not to scale.

Actually unlike in the simple case just considered, in real measurements the value of $\tau$ is not known *a priori* and several $V_g$ scans are performed before the dips amplitudes cross, the crossing point being obtained by interpolation. More over the relaxations are not exactly logarithmic (ageing effect[3]). The real situations can be simulated numerically. The value of $\tau$ then also depends on the waiting time between each $V_g$ scan (if $t_{scan}$ is not negligible compared to it), but we find that equation (1) still gives a good estimate. Thus $\tau$ is not expected to be a constant but to strongly depend on the measurement protocol, as observed experimentally.